\documentclass{aa}  
\usepackage{graphicx}
\usepackage{txfonts}
\usepackage{hyperref}
\usepackage{multirow}
\usepackage{xcolor}
\usepackage{array, booktabs, longtable}
\usepackage{makecell, multirow}
\usepackage{subcaption}
\usepackage{siunitx}
\defcitealias{Donevski2020}{DD20}
\usepackage[normalem]{ulem}
\usepackage{graphicx}
\usepackage{caption}
\usepackage{float}
\let\linenumbers\relax

\begin{document} 

\title{Dust removal timescale in galaxies across cosmic time}
\subtitle{}
\titlerunning{Dust removal timescale}

\author{A. Le\'sniewska \inst{1,2} \and J.~Hjorth \inst{1} \and C.~Gall \inst{1}}

\institute{DARK, Niels Bohr Institute, University of Copenhagen, Jagtvej 155A, DK-2200 Copenhagen N, Denmark\\
\email{aleksandra.lesniewska@nbi.ku.dk}
\and
Astronomical Observatory Institute, Faculty of Physics and Astronomy, Adam Mickiewicz University, ul. S{\l}oneczna 36, 60-286 Pozna{\'n}, Poland\\
}

\date{Received X, X; accepted X, X}

\abstract 
{Understanding the evolution of dust in galaxies is crucial because it affects the dynamics and cooling of gas, star formation, and chemical evolution. Recent work on dust removal in galaxies indicates timescales of gigayears, with old stellar populations and active galactic nuclei the primary drivers of this process. However, most statistically significant studies have focused on low redshifts, \( z < 0.4 \). Here we determine the dust removal timescale in galaxies over a wide range of redshifts, up to \( z \sim 5 \). We used publicly available catalogue data of infrared-selected galaxies observed by \textit{Herschel}. Using the inferred dust masses, stellar masses, and stellar ages, we calculated the dust removal timescale in a sample of more than 120,000 galaxies. We find that, with increasing redshift, the dust removal timescale decreases from 1.8\,Gyr at redshift \( z \sim 0.05 \) to less than 450\,Myr at \( z > 3 \). Galaxies at higher redshifts undergo more efficient dust removal than galaxies at lower redshifts, likely driven by active galactic nucleus activity, supernova shocks, and astration. These findings indicate that dust removal evolves over cosmic time, reflecting the changing mechanisms regulating the dust content of galaxies as the Universe evolves.}


\maketitle

\section{Introduction}
Dust formation in galaxies is relatively well understood. Dust is formed by supernovae (SNe), asymptotic giant branch stars, and through growth in the interstellar medium \citep[ISM; e.g.][]{Draine2009, Michalowski2010, Gall2011c}. According to \cite{Gall2018}, dust forms on a short timescale ($<$10--100\,Myr) through SNe and interstellar accretion \citep{Lesniewska2019}. Galaxies in the early Universe formed dust rapidly \citep{Langeroodi2024}. Recent observations from the \textit{James Webb} Space Telescope have revealed significant dust masses of about 10$^{8}$\,M$\odot$ at redshift $\simeq$ 8 \citep[e.g.][]{Witstok2023}. As the Universe evolved, galaxies became increasingly dust-rich and eventually reached a peak in the dust mass function, after which the late Universe became comparatively more dust-free.

Dust removal, on the other hand, is a process that remains largely unexplored and poorly described. There are several potential mechanisms contributing to ISM removal. Astration, where the ISM (including dust) is incorporated into new stars \citep[e.g.][]{Schawinski2014}, is more effective in galaxies with high star formation efficiencies. Dust can also be destroyed by stellar explosions, such as SNe \citep[e.g.][]{Barlow1978}, or heated by winds from low-mass stars during their planetary nebula phase \citep{Conroy2015}. Galactic outflows \citep{Bianchi2005} and active galactic nucleus (AGN) feedback \citep[e.g.][]{Fabian2012, Piotrowska2022} can also significantly affect the ISM.

Theoretical predictions of dust removal rates in galaxies span a huge range; the removal by SN remnants takes tens of millions of years \citep{Temim2015, Lakicevic2015}, while sputtering from SN forward shocks can take 1--3\,Gyr \citep{Gall2018}. Observational data indicate that processes responsible for the removal of the ISM from galaxies operate over a range of timescales \citep[e.g.][]{Whitaker2021,Michalowski2019}.

Observational studies have provided initial estimates of dust and gas removal timescales in galaxies at redshift \(z < 0.32\) \citep{Michalowski2019}. These studies identified an evolutionary trend for 61 \textit{Herschel}-detected dusty early-type galaxies (ETGs) in which the dust content declines exponentially with time, with a maximum dustiness of $M_{\rm dust}$/$M_{\rm stellar}$ = 10$^{-2.5}$ and a removal timescale of approximately 2.5\,Gyr. In a later study of 2,050 ETGs, \cite{Lesniewska2023} find comparable dust removal timescales of approximately 2.3\,Gyr and suggest that dust removal is driven by feedback from older stellar populations, including Type\,Ia SNe and planetary nebulae. Additionally, AGN feedback has been proposed as a potential mechanism for ISM removal in these galaxies \citep{Nadolny2024, Ryzhov2025}.

Recent observations have led to the detection of quenched galaxies, systems that have ceased forming stars,  at higher redshifts (e.g. \citealt{Whitaker2013}, \citealt{Looser2024}, \citealt{Baker2025}). Studying these early Universe galaxies, where dust removal processes likely dominate over dust formation, is crucial for understanding galaxy evolution and the subsequent transformation of their ISM. \cite{Lee2024} find that in several \( z > 1 \) galaxies, the maximum dustiness shifted to lower values, around $M_{\rm dust}$/$M_{\rm stellar}$ = 10$^{-3}$, indicating a lower dust production efficiency or a faster removal (or a selection effect, as these galaxies were chosen for their quiescent nature). High-\(z\) star-forming galaxies have shorter dust depletion times, indicating a higher star formation rate (SFR), despite higher gas fractions \citep[e.g.][]{Liu2019}. However, studies of high-\( z \) galaxies remain limited to small samples \citep[e.g.][]{Donevski2020, Lee2024}.

Here we analyse a large sample of over 120,000 \textit{Herschel}-detected galaxies to quantify dust removal timescales across a broad redshift range, up to \( z \sim 5 \). Our investigation focuses on the relationship between key physical parameters, such as the dependence of the dust-to-stellar mass ratio on the stellar age. We adopt a cosmological model with \( H_0 = 70 \) km s\(^{-1}\) Mpc\(^{-1} \), \( \Omega_{\Lambda} = 0.7 \), and \( \Omega_m = 0.3 \).

\section{Data and sample}
We used the Galaxy And Mass Assembly (GAMA) database (\citealt{Driver2011,Driver2016,Baldry2018,Smith2011}\footnote{\url{http://www.gama-survey.org}}) to investigate the physical properties of 120,061 galaxies observed, among others, by \textit{Herschel} in three GAMA fields (G09, G12, G15). Quantities such as dust mass ($M_{\rm dust}$), stellar mass ($M_{\rm stellar}$), and mass-weighted age (hereafter referred to as stellar age) were derived by \cite{Driver2016} through spectral energy distribution (SED) fitting applying 21 photometric bands using the Multi-wavelength Analysis of Galaxy Physical Properties (MAGPHYS; \citealt{daCunha2008}). The result is a uniform sample of galaxies from the GAMA DR3 with consistently characterised stellar population and ISM properties. For further analysis, we binned the sample into nine redshift bins, from 0 to 0.9, each 0.1 wide in redshift.

\citet[hereafter DD20]{Donevski2020} analysed 300 massive ($M_{\rm stellar}$ $>$ 10$^{10}$\,M$_\odot$) dusty star-forming galaxies (DSFGs), up to $z = 5.234$. These galaxies have been observed with ALMA and by \textit{Herschel}. \citetalias{Donevski2020} derived physical properties such as dust mass, stellar mass, and mass-weighted stellar ages (hereafter referred to as stellar age) through multi-band SED fitting using the Code Investigating GALaxy Emission (CIGALE; \citealt{Boquien2019}\footnote{\url{https://gitlab.lam.fr/cigale/cigale}}. The \citetalias{Donevski2020} final sample of 300 DSFGs has been selected based on a detection threshold of S/N $\geq$ 3 in a minimum of five photometric bands spanning the mid-IR to far-IR/sub-millimetre range, and S/N $\geq$ 5 in at least ten photometric bands across the optical to near-IR range. In our analysis, the sample was divided into three redshift bins: 0.9--1.21, 1.21--2.57, and 2.57--5.234.

\subsection{SED fitting effects}
Both SED fitting approaches described above assume stellar population synthesis models of \citet{Bruzual2003}, a \citet{Chabrier2003} initial mass function (IMF), and the dust attenuation model by \citet{Charlot2000}. However, the codes differ in their assumptions about star formation history (SFH) and dust emission models, both of which can influence derived physical parameters such as $M_{\rm stellar}$, $M_{\rm dust}$, and stellar ages. Several studies have investigated the extent of these differences and their impact on output parameters.

\cite{Carnall2019} analysed how four parametric SFH models; exponentially declining, delayed exponentially declining, lognormal, and double power law; affect estimates of galaxy stellar masses, star-formation rates, and mass-weighted ages obtained with SED fitting code BAGPIPES \citep{Carnall2018}. The authors show that each SFH model significantly affects the specific SFR (SFR/$M_{\rm stellar}$), which can bias the star-forming main sequence (MS) and favour younger stellar populations. Inferences of $M_{\rm stellar}$, SFR, and mass-weighted age vary by at least 0.1, 0.3, and 0.2 dex, respectively, depending on the chosen SFH model.

\cite{Hunt2019} compared SED fitting models, including MAGPHYS and CIGALE, using the KINGFISH sample (\textit{Herschel} observations) of nearby star-forming galaxies. The results of the two SED fittings were not compared with each other but to independent methods for determining individual physical parameters (see their Sect. 3.2.). Despite the differences in model assumptions (SFH and dust attenuation), both models closely matched the observed SEDs with median root-mean-square residuals of 0.05 for CIGALE and 0.06 for MAGPHYS. This indicates that the observed SED is reproduced reasonably well across all wavelengths, within the typical flux uncertainties.

\cite{Pacifici2023} compared 14 widely used SED fitting codes, including MAGPHYS and CIGALE, applied to CANDELS (Cosmic Assembly Near-infrared Deep Extragalactic Legacy Survey; \citealt{Grogin2011, Koekemoer2011}) datasets: (i) at z $\sim$ 1 with rest-frame photometry of near-UV to the near-IR, (ii) at z $\sim$ 1 with additional IR photometry, (iii) at z $\sim$ 3 photometry of rest-frame far-UV to the near-IR. To characterise the performance of each code, the authors examined the distributions of physical parameters derived from each SED fitting result. (i) The stellar mass distributions are largely consistent across codes, indicating that varying modelling assumptions have minimal influence on this parameter \citep[see also][]{Santini2015}; this suggests that $M_{\rm stellar}$ is the most reliably constrained quantity obtainable through SED fitting. The authors also examined SFR and dust attenuation, finding that their distributions peak at different values across codes. Higher SFRs are typically accompanied by higher dust attenuation, pointing to a strong age--dust degeneracy in the modelling. (ii) Additional IR photometry is valuable for better constraining dust emission, which in turn improves estimates of dust attenuation and SFR. Fits without IR photometry overestimate dust attenuation by about 0.2 dex. This effect is most evident in galaxies with moderate SFRs and dust content, where the attenuation prior can bias results towards higher values. (iii) The results closely resemble those in case (i), showing comparable stellar masses and elevated dust attenuation values at high SFRs. Emission from young stars is less age-degenerate, making it easier to separate stellar population (i.e. the SFH) and reducing code discrepancies. All codes indicate low dust attenuation, as expected for H-band-selected (1.65\,$\mu$m) galaxies at $z\sim$ 3, where dusty galaxies are typically too faint.

\cite{Paspaliaris2023} provided a direct comparison between physical parameters derived using MAGPHYS and those obtained with CIGALE, based on the GAMA sample. They found that $M_{\rm stellar}$ are consistent, with a typical scatter of around $\pm$ 0.25\,dex, and no significant systematic offset, as indicated by a median difference of only 0.0006\,dex. The dust mass estimations show increased scatter, around $\pm$ 1\,dex, with CIGALE values being on average 0.2\,dex higher than those from MAGPHYS. The discrepancy in the SED fitting results is driven by the specific dust grain models adopted in the analysis \citep[for details, see Appendix A of][]{Paspaliaris2023}. The sample from \citetalias{Donevski2020} was obtained using the dust model of \cite{Draine2014}, whereas \cite{Paspaliaris2023} employed \textsc{THEMIS} \citep{Jones2013,Jones2017,Kohler2014}. We acknowledge that the two studies adopt different dust models. On average, dust masses derived using the \cite{Draine2014} model are approximately 0.15\,dex higher than those obtained with \textsc{THEMIS}, whereas SFR and $M_{\rm stellar}$ estimates remain largely consistent between the two models \citep{Nersesian2019}.

In a recent study of nearly 300 DSFGs from the COSMOS (Cosmic Evolution Survey; \citealt{Scoville2007}) field, \cite{Gentile2024} compared the physical parameters derived from SED fitting using MAGPHYS and CIGALE. While the mean $M_{\rm stellar}$ values of the sample were comparable, more significant discrepancies were found in the mean $M_{\rm dust}$. CIGALE predicts $M_{\rm dust}$ on average 0.3\,dex higher than MAGPHYS. Despite the difference in the mean $M_{\rm dust}$ values and their measurement uncertainties, $M_{\rm dust}$ distribution coincides within the scatter, $\sigma$ of 0.5 for MAGPHYS and $\sigma$ of 0.46 for CIGALE. The mean SFRs differ by an offset of 0.12\,dex, with CIGALE yielding systematically higher values than MAGPHYS.

Based on the above works we see that $M_{\rm stellar}$ is consistent between MAGPHYS and CIGALE. To bring $M_{\rm dust}$ and SFRs from the \citetalias{Donevski2020} sample (CIGALE-based) in line with those from the GAMA sample (MAGPHYS-based), we lowered the \citetalias{Donevski2020} $M_{\rm dust}$ values by 0.3\,dex and reduced the SFRs in the \citetalias{Donevski2020} sample by 0.1\,dex. The stellar mass and mass-weighted age estimates from the original GAMA and \citetalias{Donevski2020} catalogues are used without modification. For galaxies at redshifts \( z > 0.9 \), the age estimates are limited by the age of the universe at their respective redshifts. These galaxies cannot be much older than suggested by the SED fitting, as their inferred stellar ages are constrained by the age of the Universe at the observed redshift. A decrease by 0.2\,dex in stellar age would reduce the inferred dust removal timescale to below 300\,Myr.

\subsection{Selection effects} 
Selection effects may affect our results.
Our analysis relies on infrared data from \textit{Herschel}, which detects high-\( z \) galaxies only if they contain significant dust amounts. With increasing redshift, \textit{Herschel} limits detection to the most IR-luminous galaxies, biasing the sample towards dusty, usually actively star-forming systems. This may contribute to an observed increase in the maximum dust-to-stellar mass ratio (normalisation $A$), reflecting selection effects rather than a purely physical trend.

\begin{figure*}
\centering
\includegraphics[width=\textwidth]{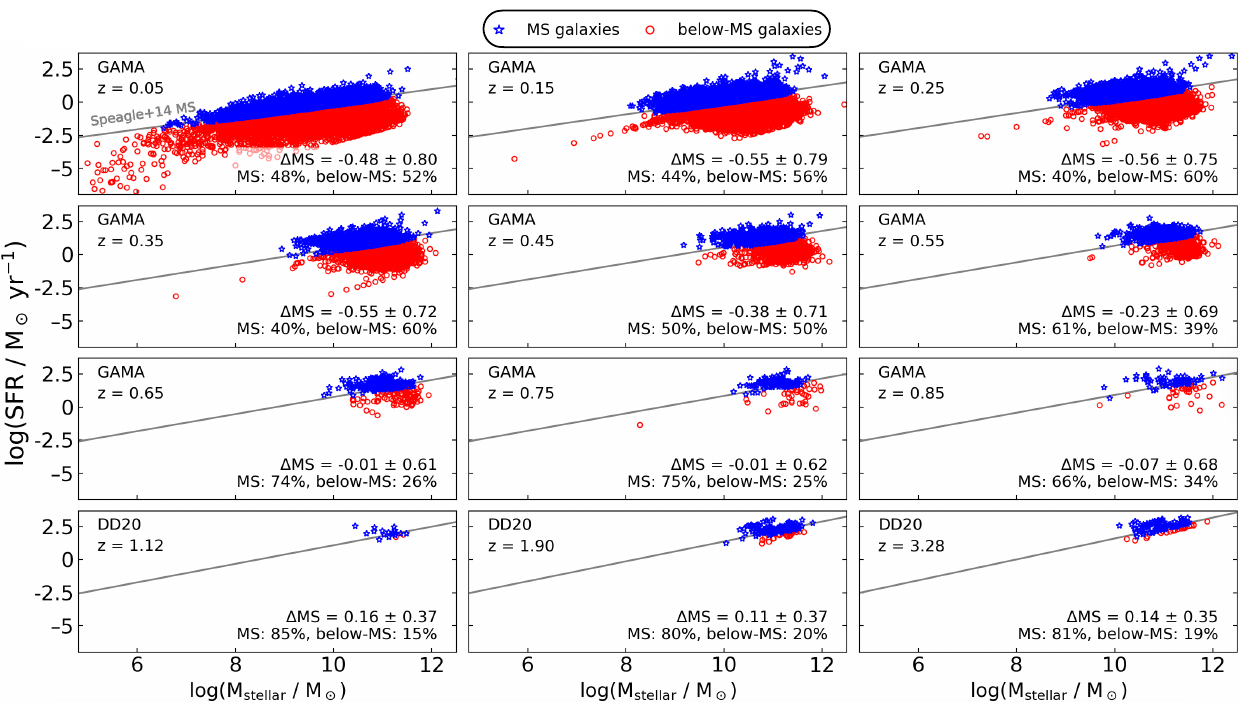}
\caption{SFR as a function of stellar mass for the GAMA sample (top three rows) and the \citetalias{Donevski2020} sample (bottom row). The star-forming MS at median redshift of each redshift bin (grey line) based on \cite{Speagle2014} is shown. Colour coding distinguishes galaxies on or above the MS (blue stars) and galaxies below the MS (red circles). The MS offset of each redshift bin is presented at the bottom right of each panel together with the fraction of MS and below-MS galaxies.}
\label{MS-panels}
\end{figure*}

To evaluate whether GAMA and \citetalias{Donevski2020} trace comparable populations, we compared their locations in the SFR-stellar mass plane (Fig.\,\ref{MS-panels}).We implemented the redshift-dependent MS from \cite[][Eq.\,28]{Speagle2014}, which reliably describes galaxies out to \(z \sim 5\), to distinguish between galaxy populations: those more than 0.2\,dex below the MS (red circles), while galaxies on or above it are considered MS galaxies (blue stars). GAMA transitions from a mixed population at low redshift (48\% on the MS at $z \approx 0.05$) to predominantly MS galaxies at $z \approx 0.85$ (66\%). The \citetalias{Donevski2020} sample is consistently MS-dominated, with 81--85\% of galaxies above the threshold. These results confirm that both samples predominantly represent typical star-forming galaxies, and the observed dust removal evolution reflects real physical changes rather than selection bias.

\section{Analysis and results}
The selected data were used to calculate the dust removal timescale and the maximum dustiness that galaxies can reach. To obtain the timescale of dust removal, we implemented an exponential relation between the dust-to-stellar mass ratio and stellar population age established by \cite{Michalowski2019}, described as

\begin{equation}\label{MM19function}
    \frac{M_{dust}}{M_{stellar}} = A \cdot e^{-age/\tau},
\end{equation}

\noindent
where \( A \) is the normalisation constant, \( age \) is the stellar population age, and $\tau$ is the dust removal timescale. 

\begin{figure}[t]
\includegraphics[width=0.5\textwidth]{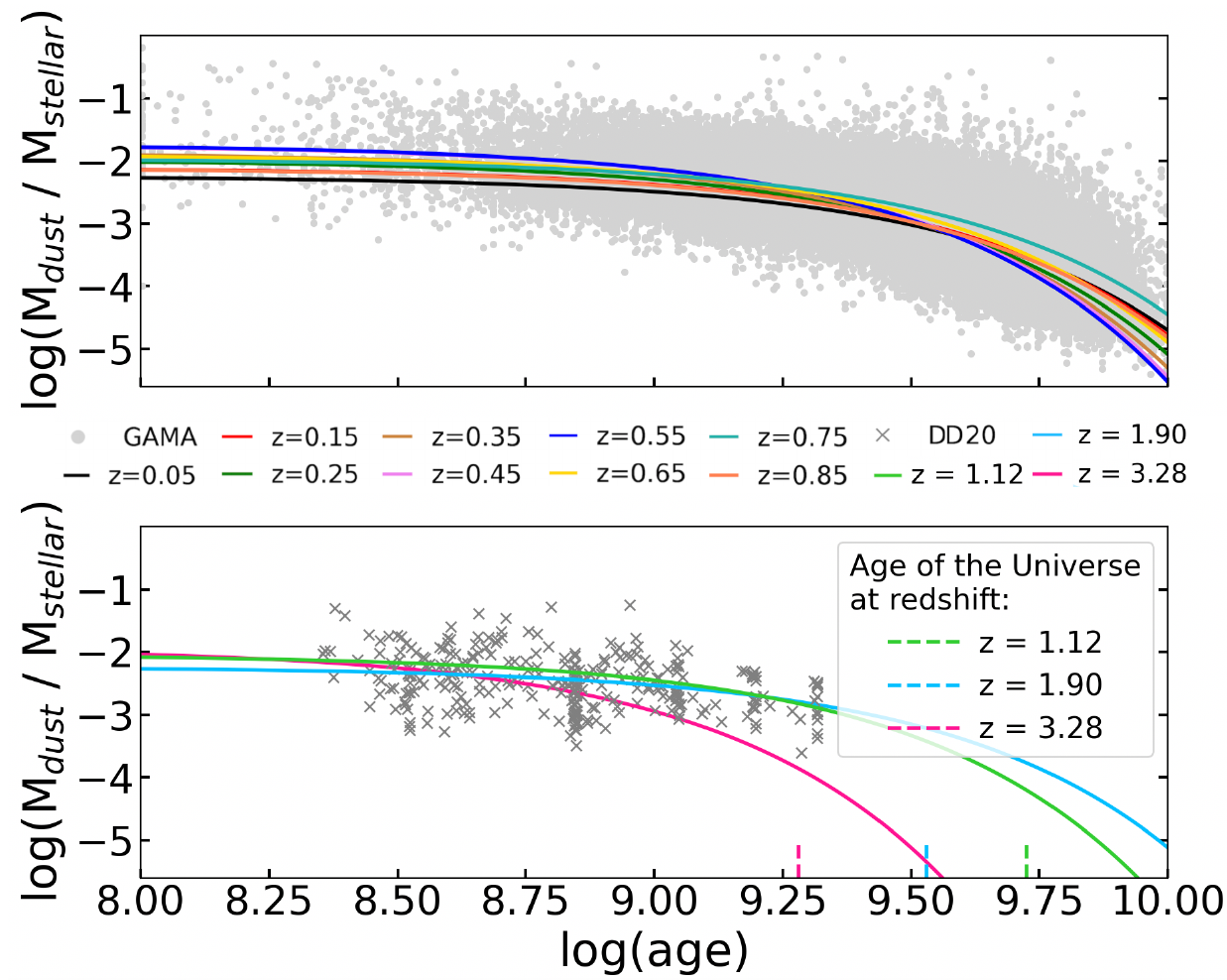}
\caption{Dust-to-stellar mass ratio as a function of stellar age. Top: GAMA galaxies (light grey dots) and exponential fits for redshift bins up to 0.9. Bottom: \citetalias{Donevski2020} galaxies (grey cross markers) and exponential fits for redshift bins from \( z = 0.9\) to \( z \sim 5\). Vertical dashed lines represent the age of the Universe at the corresponding redshift (colours correspond to the redshift bins). Plots for the individual redshift bins and fits are included in Appendix\,\ref{Removal-panels}.}
\label{MdMs}
\end{figure}

\begin{figure}
\includegraphics[width=0.5\textwidth]{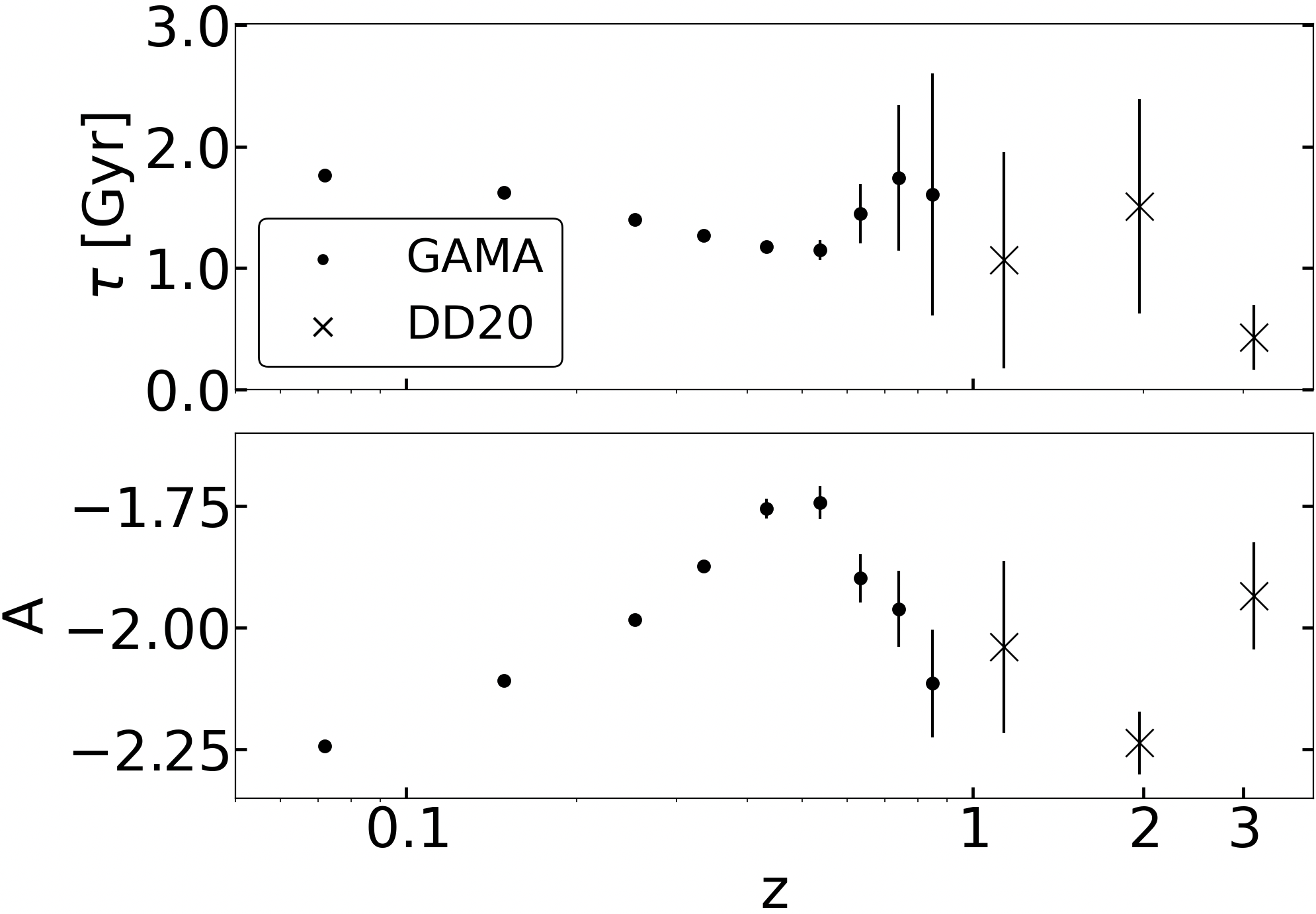}
\caption{Top: Dust removal timescale as a function of redshift. Bottom: Normalisation constant as a function of redshift.}
\label{Tau}
\end{figure}

The numerical results of our analysis are presented in Table\,\ref{Table}. Figure \ref{MdMs} (top) shows the dust-to-stellar mass ratio versus stellar age, with exponential fits to the GAMA sample within redshift bins. The bottom panel presents fits to \citetalias{Donevski2020}. Figure \ref{Tau} (top) presents the dust removal timescale, decreasing from 1.77$\pm$0.04\,Gyr at \( z\sim0.05 \) to 0.43$\pm$0.33\,Gyr at \( z\sim3.28 \), with larger uncertainties for \( z>0.9 \). The normalisation constant, associated with the maximum dustiness of galaxies, increases with redshift (Fig.\,\ref{Tau}, bottom). Galaxies at redshift $z\sim0.05$ have approximately $M_{\rm dust}$/$M_{\rm stellar}$ = 10$^{-2.24}$. On the other hand, there is $M_{\rm dust}$/$M_{\rm stellar}$ = 10$^{-1.93}$ in galaxies at redshift $z > $ 3. Young galaxies (\(\log_{10} (\text{age}) \leq 8.5\)) in the GAMA sample are not well described by the exponentials, which underestimates the dustiness ratio, indicating higher true values. Plots for the individual redshift bins and exponential fits are included in Appendix\,\ref{AppendixRemoval-panels}.

\begin{table}[ht]
\centering
\caption{Medians of the dust removal timescale ($\tau$) in different redshift bins and the normalisation constant from fitting exponential functions to the data of Fig. \ref{MdMs}.}
\begin{tabular}{c|c|c|c}
 \hline
 \( z \) & \( \tau \) [Gyr] & \(A\) & \# of galaxies \\

 \hline
 \hline
  \multicolumn{4}{c}{GAMA} \\  
 $\leq$ 0.1 & 1.77 $\pm$ 0.04 & $-2.24 \pm 0.01$ & 16,649 \\ 
 0.1 -- 0.2 & 1.63 $\pm$ 0.02 & $-2.11 \pm 0.01$ & 44,819 \\ 
 0.2 -- 0.3 & 1.40 $\pm$ 0.02 & $-1.98 \pm 0.01$ & 34,089 \\ 
 0.3 -- 0.4 & 1.27 $\pm$ 0.03 & $-1.87 \pm 0.01$ & 17,452 \\ 
 0.4 -- 0.5 & 1.18 $\pm$ 0.05 & $-1.75 \pm 0.02$ & 4,932 \\ 
 0.5 -- 0.6 & 1.15 $\pm$ 0.09 & $-1.74 \pm 0.03$ & 1,450 \\
 0.6 -- 0.7 & 1.45 $\pm$ 0.25 & $-1.90 \pm 0.05$ & 417 \\ 
 0.7 -- 0.8 & 1.74 $\pm$ 0.67 & $-1.96 \pm 0.08$ & 155 \\ 
 0.8 -- 0.9 & 1.61 $\pm$ 1.12 & $-2.11 \pm 0.11$ & 98 \\ 
 \hline
 \multicolumn{4}{c}{\citetalias{Donevski2020}} \\  
 0.9 -- 1.21 & 1.07 $\pm$ 0.93 & $-2.04 \pm 0.17$ & 20 \\ 
 1.21 -- 2.57 & 1.51 $\pm$ 0.87 & $-2.24 \pm 0.06$ & 153 \\ 
 2.57 -- 5.234 & 0.43 $\pm$ 0.33 & $-1.93 \pm 0.11$ & 124 \\ 
 \hline
\end{tabular}
\label{Table}
\end{table}

To verify the robustness of our GAMA galaxy sample, we studied five subsamples of galaxies. These confirm the observed trends, where the dust removal timescale decreases with increasing redshift. Similarly, the normalisation constant follows the same pattern as in the main sample, increasing with redshift. The results of this evaluation are presented in Appendix \ref{AppendixGAMA}.

\section{Discussion}
Our analysis shows that the dust removal timescale decreases with redshift, ranging from approximately 1.8\,Gyr at \( z < 1 \) to about 0.4\,Gyr at \( z > 3 \). This suggests that more efficient mechanisms dominate the dust-clearing process at earlier epochs. Additionally, we find that the normalisation constant increases with redshift, indicating that younger galaxies at earlier cosmic epochs are dustier.

\subsection{Drivers of dust removal}
The most efficient processes for dust removal in galaxies can be narrowed down to four main physical mechanisms: (i) astration, (ii) the destruction or (iii) heating of dust grains, and (iv) the ejection of dust from galaxies without complete destruction \citep[e.g.][]{Veilleux2021}. While several detailed mechanisms for low-\(z\) galaxies have been discussed in previous studies \citep{Michalowski2024}, we focus here on these four key processes that are particularly relevant to our analysis.

\subsubsection{Astration}
Astration is closely tied to the process of star formation. The process can be accelerated by frequent galaxy mergers in high-density environments \citep[e.g.][]{Lin2010}. Elevated SFRs at high-\(z\) have been observed \citep[e.g.][]{Reeves2024, Yuan2024}.
However, as galaxies get older, the SFR declines, leading to a corresponding decrease in the effectiveness of the astration mechanism. Astration can extend beyond a few hundred million years, as it gradually depletes gas and dust through sustained star formation. 
For the galaxies in our sample at \( z > 0.9 \), which were selected as star-forming galaxies, astration likely contributes substantially to the dust removal.

\subsubsection{SN shock dust destruction}
Supernova shocks heat and accelerate the gas, enhancing dust-gas collisions and returning grain material to the gas phase as metals. This process is called sputtering and can effectively destroy dust grains.
The destruction of dust by SNe, like astration, is closely linked to the star formation activity of a galaxy. The efficiency of massive star formation, and consequently the rate of their explosions impacting the ISM, depends on the galaxy's IMF. Galaxies with a top-heavy IMF, for instance, will produce a greater number of massive stars.
The efficiency of dust destruction by SN shocks will impact the evolution of the ISM, particularly as they slow down or stop the formation of massive stars.

\subsubsection{AGN feedback}
Active-galactic-nucleus-driven outflows effectively remove or heat the ISM, especially at high redshifts where AGNs are more common due to gas inflows and mergers \citep[e.g.][]{Hopkins2006}. The AGN fraction increases as galaxy merging progresses \citep[e.g.][]{Ellison2011}. \citet{Comerford2024} observed an excess of AGNs in mergers, relative to non-mergers, of 1.8 in major and 1.5 in minor mergers from MaNGA (Mapping Nearby Galaxies at Apache Point Observatory; \citealt{Bundy2015, Law2015}) data, with similar results from SDSS (Sloan Digital Sky Survey) and GAMA samples \citep[e.g.][]{Driver2011, Gao2020}. \citet{Pantoni2021} analysed 11 DSFGs at \(z \sim 2\) detected in the (sub-) millimetre regime and identified AGN feedback signatures in four of them based on X-ray and radio data. 
  
Active galactic nucleus duty cycles (periods of AGN activity) typically are much less than 1 Gyr \citep[e.g.][]{Novak2011}, with outflows operating on timescales of a few hundred million years \citep{Hardcastle2018}, aligning with the 0.4\,Gyr dust removal timescale in our high-\(z\) sample. Observationally, \citet{Singh2024} found quiescent galaxies at \(z \sim 3.1\) where quenching timescales remain under 500\,Myr due to AGN feedback and morphological quenching. \citet{Sato2024} reported dust removal timescales of 0.2–0.7\,Gyr in quiescent galaxies at $z = 3$--4, mainly driven by AGN activity, supporting our finding of shorter dust removal timescales with redshift.

\subsection{Implications for galaxy evolution}
Low-\(z\) galaxies likely evolve from high-\(z\) galaxies, transitioning from spirals to ellipticals via mergers \citep{Hopkins2009}. Mergers drive supermassive black hole growth, crucial to galaxy evolution \citep[e.g.][]{Ellison2019}. Each low-\(z\) galaxy reflects its current evolutionary stage, shaped by past events. \cite{Nadolny2024} show that galaxies at \(z \sim 0.1\) may have undergone mergers, boosting star formation, while AGN activity quenches it and expels the ISM \citep[e.g.][]{Croton2006}.

The increasing normalisation constant with redshift suggests that young \( z \sim 3 \) galaxies form dust more efficiently or there is less time for dust destruction than low-\(z\) counterparts. \cite{Algera2025} found $M_{\rm dust}$/$M_{\rm stellar}$ of \( 10^{-2.7} \) to \( 10^{-1.8} \) in \( 6.5 \leq z \leq 7.7 \) galaxies, consistent with our results. The increase in normalisation may reflect the evolution of star formation density, which peaks at \( z \sim 2 \) \citep{Madau2014ARA&A}. Observations from UV to IR show a steady star formation increase from early epochs, peaking at \( z \sim 2 \) before declining, linking dust accretion to star formation efficiency through gas availability, feedback, and AGN activity.

\citet{Gall2011a, Gall2011b} explore core-collapse supernovae (CCSNe) as primary dust producers and the IMF's impact on dust yield in the early Universe. A top-heavy IMF increases the CCSN rate, boosting dust production, while a Salpeter \citep{Salpeter1955} or Chabrier \citep{Chabrier2003} IMF results in lower yields. Given that the youngest galaxies studied here have $M_{\rm dust}$/$M_{\rm stellar}$ $\sim 10^{-2.41}$ to $10^{-1.63}$, this suggests efficient early dust buildup, possibly from a high-mass-biased IMF. This aligns with \citet{Gall2011a, Gall2011b}, where high CCSN rates enhance dust content. Comparing our results with their models can help constrain the IMF and dust production mechanisms, refining our understanding of early galaxy evolution.

\section{Conclusions}

In this work we examined dust removal timescales and the dust-to-stellar mass ratio across a wide redshift range, up to \( z < 5.3 \), using a large sample of galaxies from the GAMA survey and DSFGs from the COSMOS field. Our analysis reveals a clear trend of decreasing dust removal timescales with increasing redshift, from approximately 1.8\,Gyr at \( z < 0.1 \) to 0.4 Gyr at \( z > 3 \). This suggests that dust removal mechanisms are more efficient in high-z galaxies, potentially driven by enhanced star formation, CCSNe, and AGN activity.

Additionally, the increasing normalisation constant ($M_{\rm dust}$/$M_{\rm stellar}$ maximum) with redshift implies more efficient dust production in younger galaxies, supporting the notion that high-z galaxies form dust at a higher rate than their low-z counterparts. The correlation between dust accumulation and high SFRs, particularly in galaxies at high redshifts, highlights the close link between intense star formation and active dust enrichment in the early Universe. These findings align with previous works on CCSNe and high SFRs in early dust production.

Our findings suggest that AGN feedback plays a key role in setting an upper limit on the dust content of galaxies, preventing them from becoming overly dust-enriched at early epochs. This has important implications for models of dust evolution, particularly those aimed at replicating dust-to-gas and dust-to-stellar mass ratios across cosmic time. Although detection biases may influence the observed sample, especially for star-forming and starburst galaxies at high redshifts, the trends identified in this study provide valuable constraints on the mechanisms behind dust removal in galaxies across cosmic time.

\begin{acknowledgements}
We wish to thank the referee for the detailed comments which helped us to clarify our work.
We would like to thank Darko Donevski for providing data and Michał Michałowski and Danial Langeroodi for their helpful comments.

This work was supported by research grants (VIL16599, VIL54489) from VILLUM FONDEN.
CG acknowledges support from research grants (VIL25501 and VIL69896) from VILLUM FONDEN. 

GAMA is a joint European-Australasian project based around a spectroscopic campaign using the Anglo-Australian Telescope. The GAMA input catalog is based on data taken from the Sloan Digital Sky Survey and the UKIRT Infrared Deep Sky Survey. Complementary imaging of the GAMA regions is being obtained by a number of independent survey programmes including GALEX MIS, VST KiDS, VISTA VIKING, WISE, Herschel-ATLAS, GMRT, and ASKAP providing UV to radio coverage. GAMA is funded by the STFC (UK), the ARC (Australia), the AAO, and the participating institutions. The GAMA website is http://www.gama-survey.org/. This research has made use of the NASA’s Astrophysics Data System Bibliographic Services.
\end{acknowledgements}

\bibliographystyle{aasjournal}
\bibliography{main}

\begin{appendix} 
\onecolumn
\section{Dust-to-stellar mass ratio as a function of stellar age}\label{AppendixRemoval-panels}

\begin{figure*}[ht]
\includegraphics[width=\textwidth]{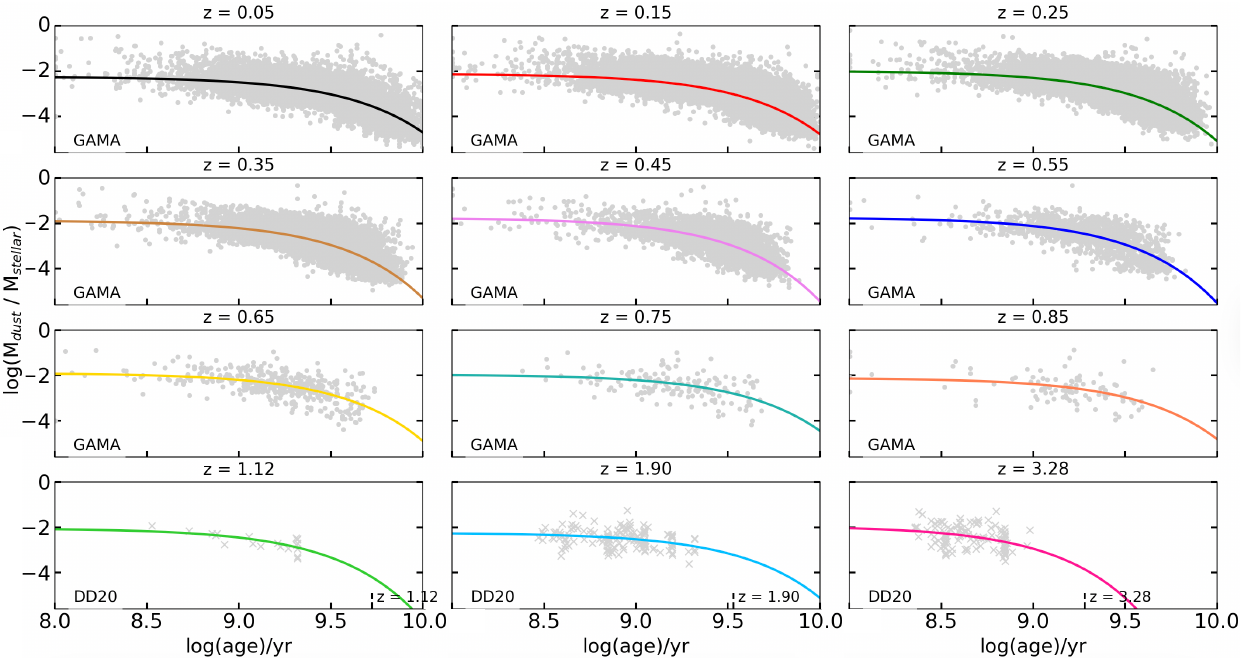}
\caption{Dust-to-stellar mass ratio as a function of stellar age and exponential fits for individual redshift bins (coloured lines). The first three rows present GAMA redshift bins and the last row presents \citetalias{Donevski2020} redshift bins. The vertical dashed black lines in the last row represent the age of the Universe at the corresponding redshift.}
\label{Removal-panels}
\end{figure*}

The dust-to-stellar mass ratio as a function of stellar age and exponential fits for GAMA and \citetalias{Donevski2020} individual redshift bins is presented in Fig.\,\ref{Removal-panels}.

\section{GAMA sample evaluation}\label{AppendixGAMA}
To verify the robustness of our GAMA galaxy sample, we studied five subsamples of galaxies. Large uncertainties of the dust removal timescale, $\tau$,  result from the `flat' decline of galaxies in the dust-to-stellar mass ratio versus\ stellar age, a small number of galaxies in a redshift bin, and/or a scatter of these galaxies, preventing the determination of the physical values of $\tau$ uncertainties.

\subsection{GAMA dusty galaxies} 
Selecting dust detection confirmed using \textit{Herschel} SPIRE $250\,\mu m$ (S/N $\geq$ 3) resulted in 35,221 galaxies.

Dusty galaxies exhibit longer dust removal timescales for redshifts  \( z < 0.4 \), reaching almost 2.5\,Gyr. Similar findings on dust and gas removal in observed and simulated ETGs at \( z \sim 0.32 \) have been reported by \cite{Michalowski2019, Lesniewska2023, Nadolny2024}. However, their analysis of the redshift dependence of dust mass removal did not lead to the conclusions presented in this study, as they found no redshift-timescale correlation. In all cases, the limited redshift range and sample size may have prevented the detection of the trend observed here. The results are presented in Fig.\,\ref{Tau-Dusty}.

\subsection{GAMA dusty spirals} 
The selections S/N $\geq$ 3 and  $r$-band S\'ersic indices \citep{Sersic1963} lower than or equal to 2.5 resulted in 28,027 galaxies.

A parameter used to describe a galaxy's morphology is the $r$-band S\'ersic index \citep{Sersic1963}, which reflects the distribution of light within a galaxy. A commonly adopted conservative limit for the S\'ersic index is $n = 2.5$ \citep[e.g.][]{Lange2015}, below which galaxies are classified as spiral galaxies. It is important to note, however, that at redshifts \( z > 0.4 \), the S\'ersic selection may also include bulge-dominated or compact late-type galaxies. The results are presented in Fig.\,\ref{Tau-Spirals}.

\subsection{GAMA massive dusty spirals}
The selections S/N $\geq$ 3, S\'ersic\, index $\leq$ 2.5, and M$_{stellar} > 10^{10}$\,M$_\odot$ resulted in 23,120 galaxies.

\citetalias{Donevski2020} analysed massive galaxies with stellar masses M$_{stellar} > 10^{10}$\,M$_\odot$. GAMA massive dusty spirals show dust removal timescales approaching 3.5\,Gyr at redshift \( z < 0.1 \). This finding highlights the prolonged dust evolution in these galaxies, suggesting that the dust content is subject not only to the removal mechanisms but also to the dust production mechanism. The results are presented in Fig.\,\ref{Tau-Massive}.

\subsection{GAMA massive DSFGs}
The selections S/N $\geq$ 3, SFR $> 10^{1.5}$\,M$_\odot$ yr$^{-1}$, M$_{dust} > 10^{8}$\,M$_\odot$, and M$_{stellar} > 10^{10}$\,M$_\odot$ resulted in 1,332 galaxies.

The massive DSFGs in \citetalias{Donevski2020} fit the criteria outlined above, which is why we also selected these galaxies from GAMA. Due to the applied parameter limits, the resulting sample consists of 1,332 galaxies, and the uncertainties obtained are the largest calculated here. The first GAMA redshift bin (\( z < 0.1 \)) contains only one galaxy, which made it impossible to obtain any fit. The results are presented in Fig.\,\ref{Tau-DustMassiveSFG}.

\subsection{GAMA massive dusty ETGs}
The selections S/N $\geq$ 3, M$_{stellar} > 10^{10}$\,M$_\odot$, and S\'ersic\, index $>$\,4 resulted in 2,778 galaxies.

\begin{figure}[h]
    \centering
    \begin{minipage}[t]{0.32\textwidth}
        \centering
        \includegraphics[width=\linewidth]{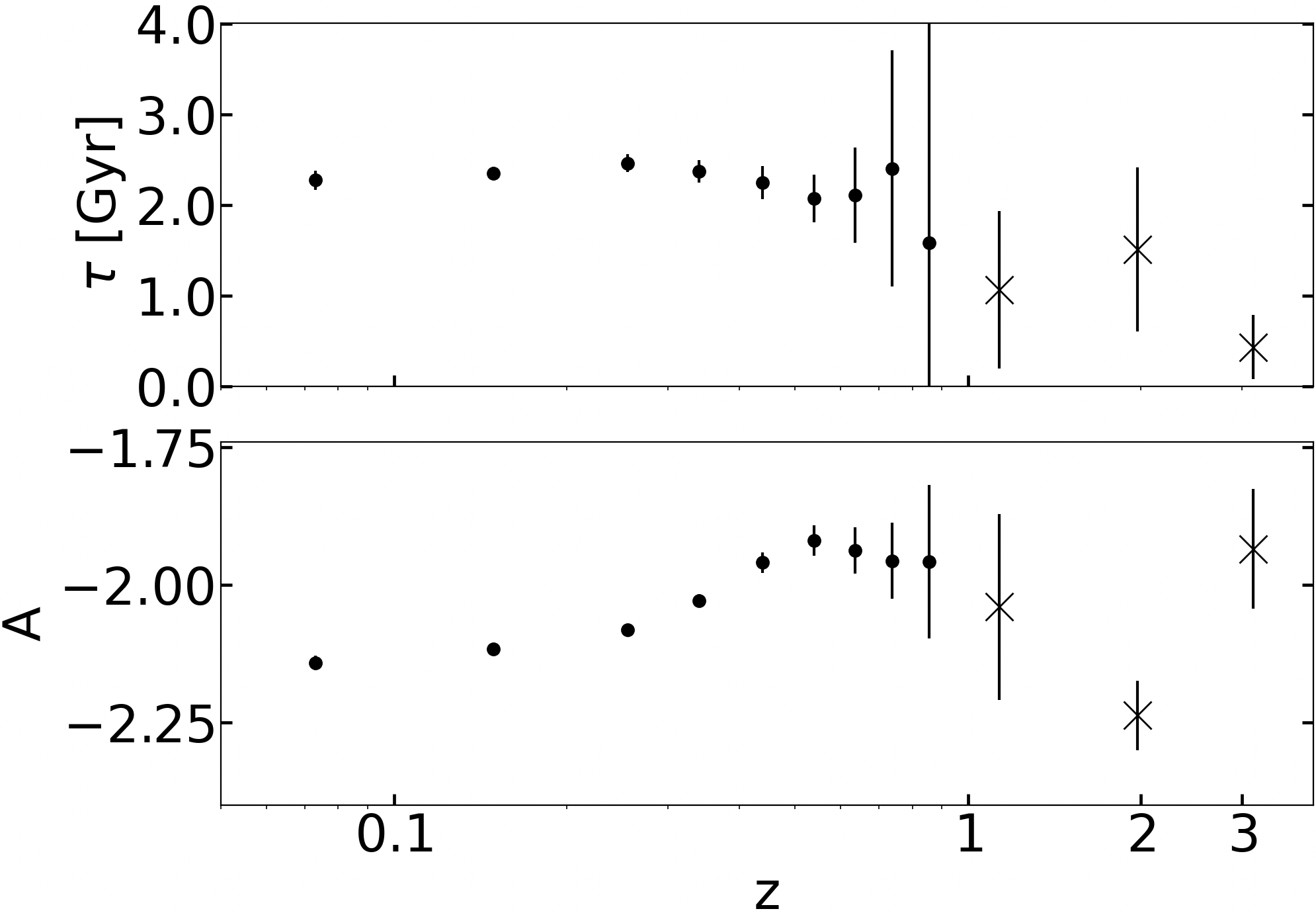}
        \subcaption{}
        \label{Tau-Dusty}
    \end{minipage}
    \hfill
    \begin{minipage}[t]{0.32\textwidth}
        \centering
        \includegraphics[width=\linewidth]{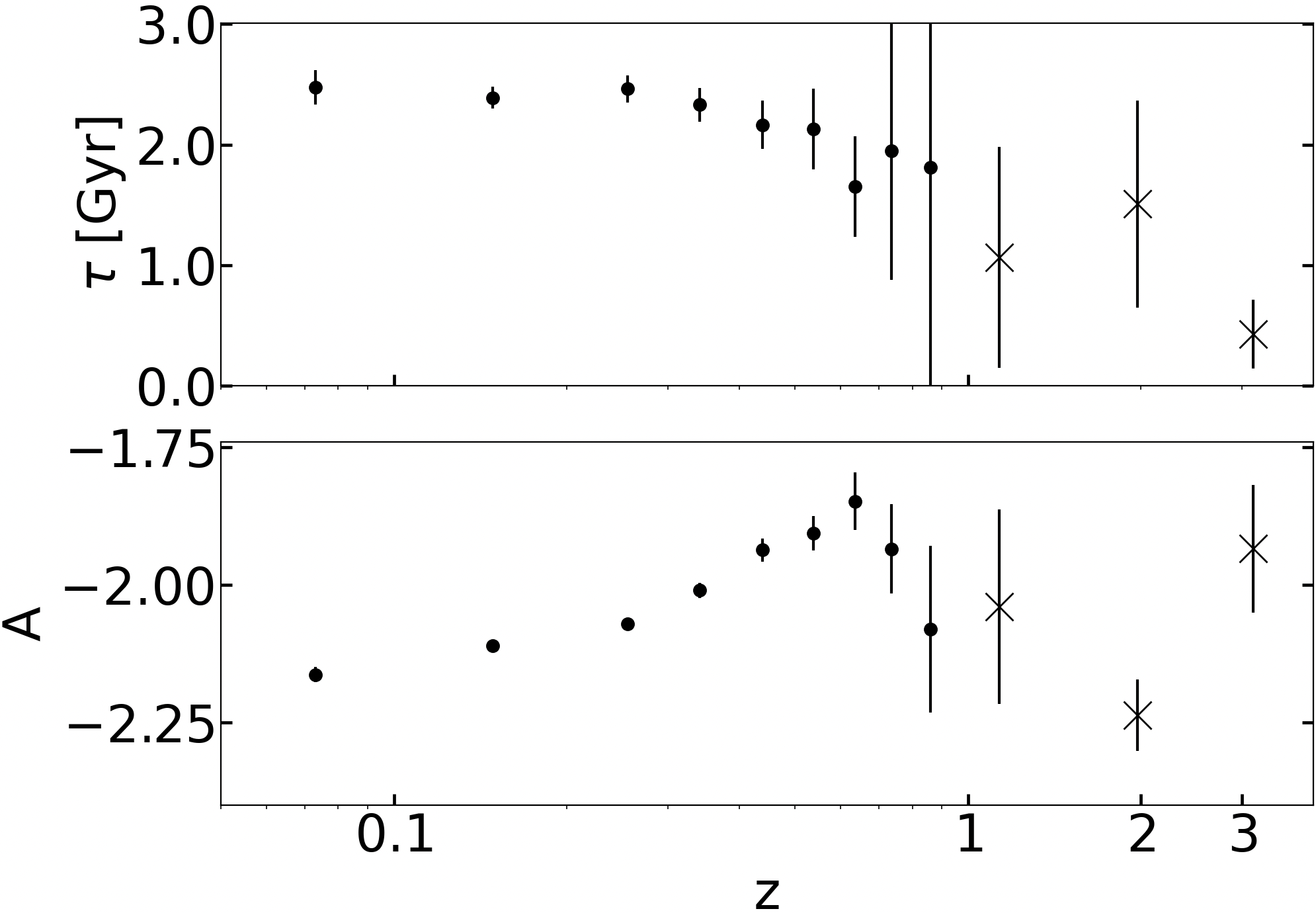}
        \subcaption{}
        \label{Tau-Spirals}
    \end{minipage}
    \hfill
    \begin{minipage}[t]{0.32\textwidth}
        \centering
        \includegraphics[height=4.2cm]{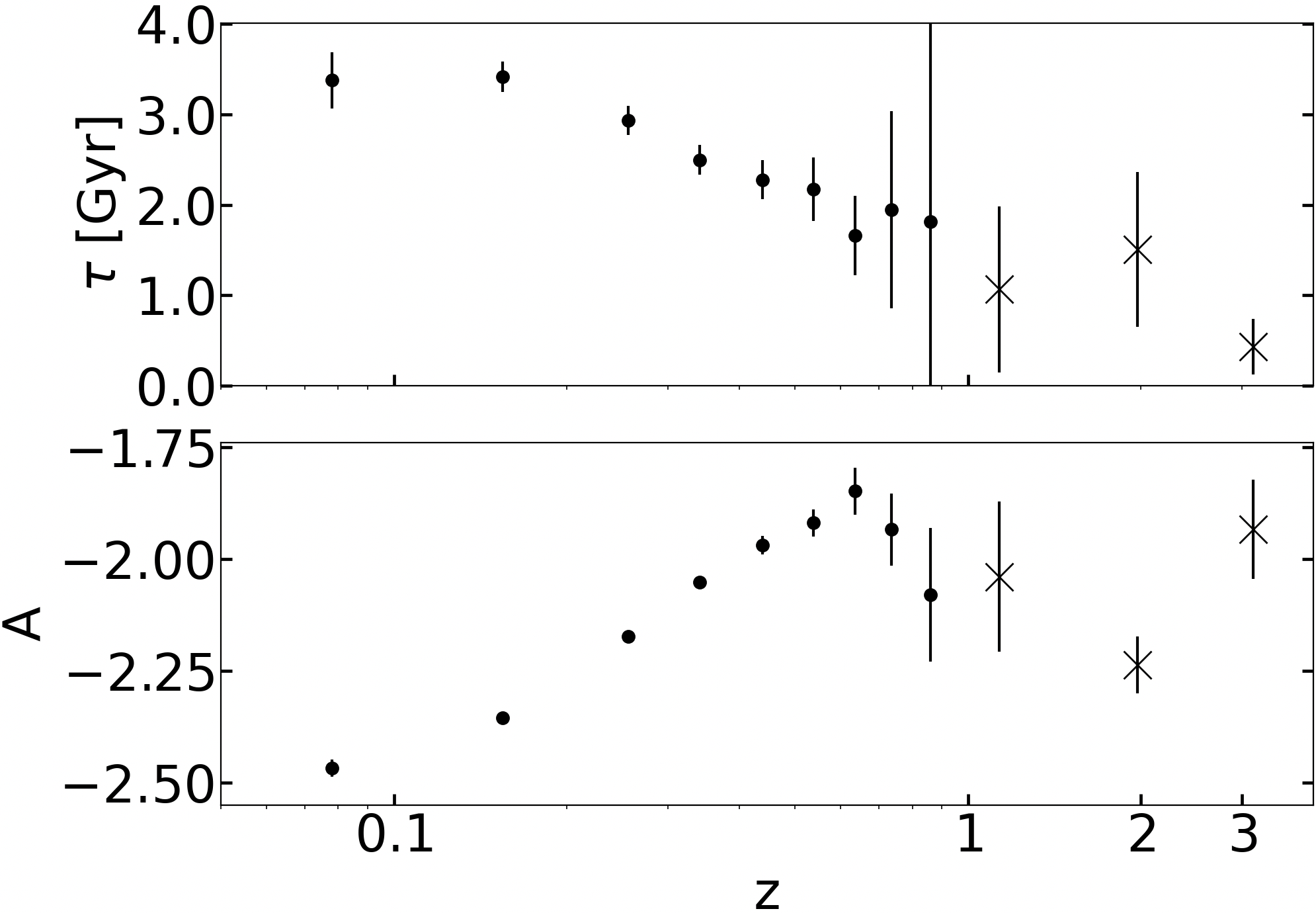}
        \subcaption{}
        \label{Tau-Massive}
    \end{minipage}

    \begin{center}
    \begin{minipage}[t]{0.32\textwidth}
    \centering
    \includegraphics[height=4.2cm]{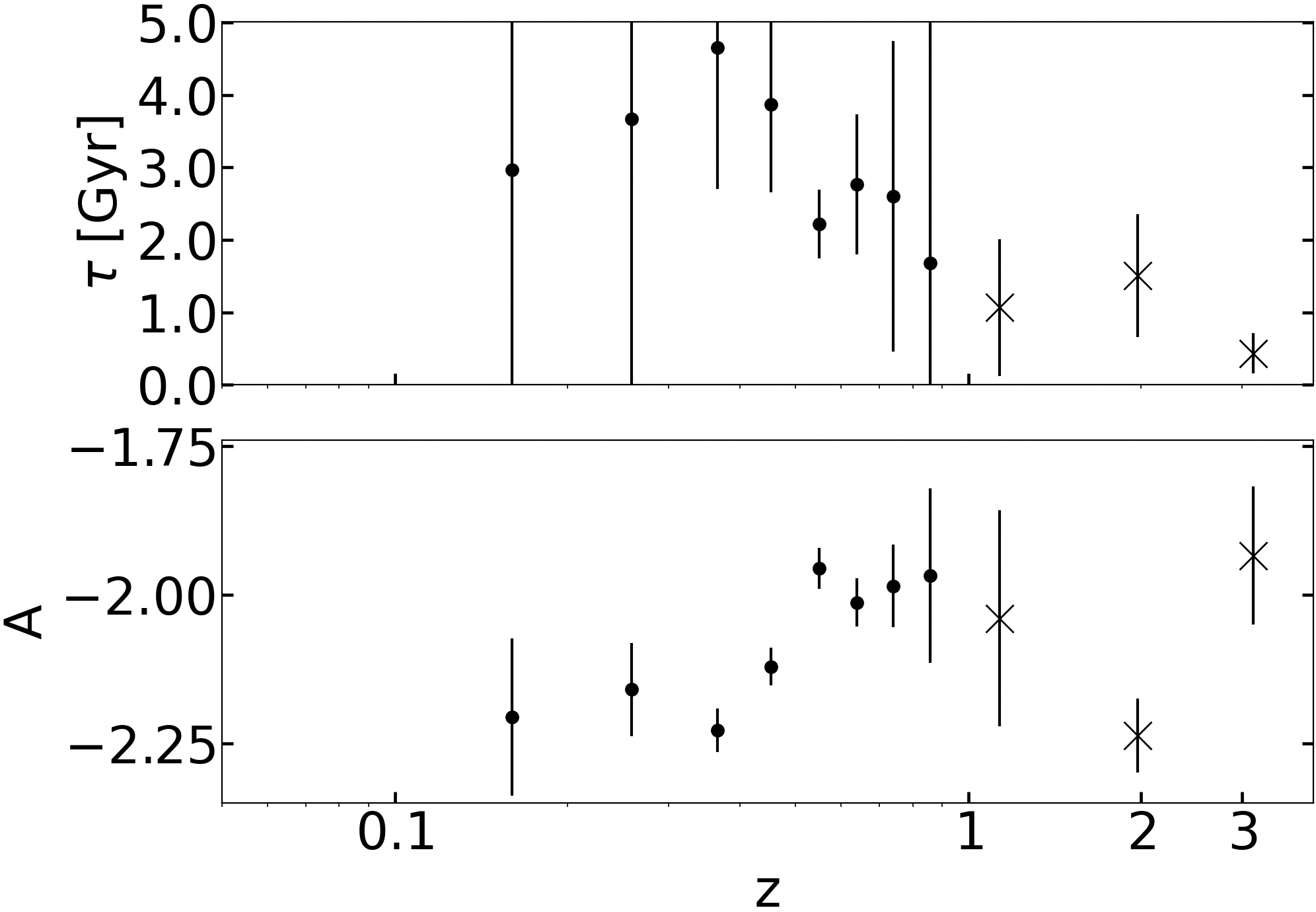}
    \subcaption{}
    \label{Tau-DustMassiveSFG}
    \end{minipage}
    \hspace{0.06\textwidth}
    \begin{minipage}[t]{0.32\textwidth}
    \centering
    \includegraphics[height=4.2cm]{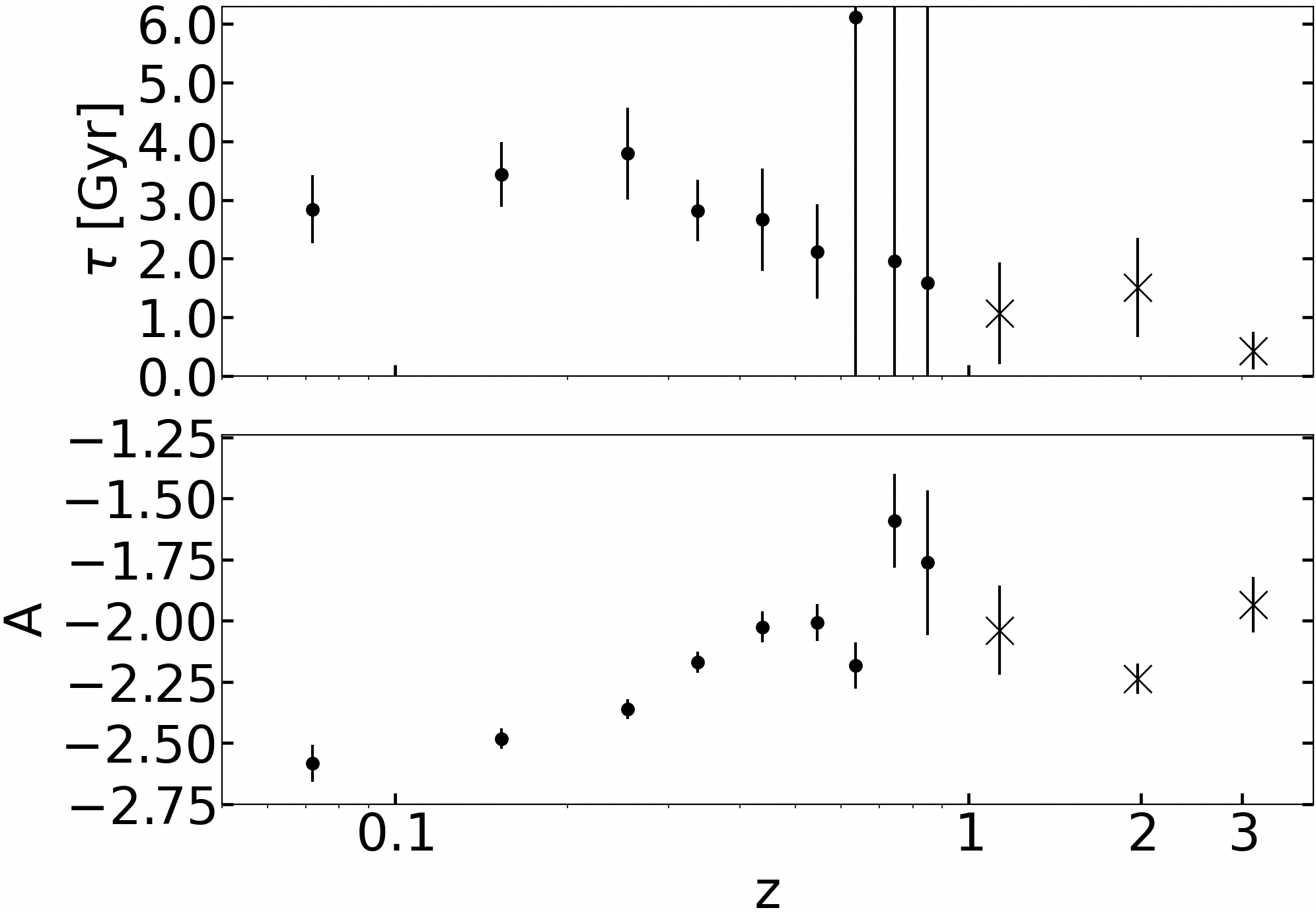}
    \subcaption{}
    \label{Tau-DustMassiveETGs}
    \end{minipage}
\end{center}

    \caption{Overview of dust removal timescales and normalisation constants for different GAMA galaxy subsamples, compared to \citetalias{Donevski2020}. Panel (a): GAMA dusty galaxies. Top: Dust removal timescale as a function of redshift for GAMA dusty galaxies (\textit{Herschel} SPIRE $250\,\mu$m S/N $\geq$ 3, redshift up to 0.9) and galaxies from \citetalias{Donevski2020} (redshift from 0.9 to 5.234). Bottom: Normalisation constant as a function of redshift. Panel (b): GAMA dusty spiral galaxies + \citetalias{Donevski2020}. Panel (c): GAMA massive dusty spiral galaxies +\,\citetalias{Donevski2020}. Panel (d): GAMA massive DSFGs + \citetalias{Donevski2020}. The GAMA redshift bin with $z < 0.1$ contains only a single galaxy, which made it impossible to obtain any fit. Panel (e): GAMA massive dusty ETGs + \citetalias{Donevski2020}. }
    \label{fig:Tau-All}
\end{figure}

The subsample of massive dusty ETGs, similar to the other subsamples and the main sample, shows a decline in the dust removal timescale. The results are presented in Fig.\,\ref{Tau-DustMassiveETGs}.

\end{appendix}
\end{document}